# On the factors affecting the high temperature insulator-metal transition in rare-earth manganites

Dipten Bhattacharya*, Pintu Das, A. Pandey, and A.K. Raychaudhuri†
National Physical Laboratory, New Delhi 110 012, India

Amitava Chakraborty
Electroceramics Division, Central Glass & Ceramic Research Institute, Calcutta 700 032, India

V.N. Ojha
National Physical Laboratory, New Delhi 110 012, India

**Abstract.** The measurement of resistivity across a wide temperature range – from 15 to 1473 K – in rare-earth manganite series of compounds reveals a very interesting feature : normally observed insulating pattern beyond $T_c$ (Curie Point) gives way to a reentrant metallic pattern around a characteristic temperature $T^*$. The transport activation barrier $E_a$ collapses to zero around $T^*$. $T^*$ is found to be dependent on the carrier concentration or the concentration of the Jahn-Teller-active $Mn^{3+}$ ions as well as on the average A-site radius ($\langle r_A \rangle$) for a fixed carrier concentration. These factors govern the effective lattice distortion and hence lead to the variation in the conduction bandwidth. Our data cover a wide range – from $T^* >> T_c$ for smaller bandwidth to $T^* \rightarrow T_c$ for larger bandwidth. These result seem to provide evidence of the onset of lattice distortion at high temperature (around $T^*$) and its variation. Since lattice distortion governs the magnetic, transport and other important behaviors significantly, our data assume importance as they offer a new measure of the effective distortion and its tunability.

PACS Nos. 75.30.Vn, 71.38.+i, 71.30.+h

In recent years the physics of rare-earth manganites (with general chemical formula $Re_{1-x}Ae_xMnO_3$; Re = La, Nd, Pr, etc. and Ae = Sr, Ca, Ba etc.) has attracted considerable interest as this series of compounds displays a number of fascinating phenomena like colossal magnetoresistance (CMR), charge ordering (CO), magnetic field driven insulator-metal transition, electronic phase segregation etc. across the entire phase diagram. A fine interplay of charge, spin and orbital effects is responsible for most of the observed phenomena. A very crucial issue in the physics of manganites is the onset of lattice distortions at high temperature; one such distortion is the cooperative Jahn-Teller (JT) effect around the $Mn^{3+}$ ions in $Mn^{3+}O_6$ octahedra. The onset of lattice distortion at high temperatures (typically at T>500 K) is believed to be the cause of the insulating state that occurs in these oxides in the paramagnetic state. It has been observed that the parent compound $LaMnO_3$ undergoes a cooperative transition associated with removal of JT distortion around the $Mn^{3+}$ ions at $T \approx 750$ K (referred to as $T_{JT}$) as seen through neutron diffraction [1]. A recent transport measurement in single crystals of $LaMnO_3$ shows that at the same temperature the resistivity (ρ) drops by nearly two orders on heating through $T_{JT}$ and it becomes temperature independent beyond $T_{JT}$ [2]. The magnetization data indicate discontinuous rise in Weiss constant which signifies concomitant rise in ferromagnetic interactions in Curie-Weiss paramagnetism beyond $T_{JT}$. It is thus clear that, at least, in the parent compound $LaMnO_3$, for $T>T_{JT}$, as the orbital order due to JT effect is removed, one does get a phase with sharply reduced resistivity. This certainly marks the onset of a qualitatively new phase at $T_{JT}$. In the absence of a better description we call this a high temperature "reentrant metallic phase".

Past experimental works on bivalent metal ion substituted systems like $Re_{1-x}Ae_xMnO_3$ (where Re changes from La to Ho and Ae = Ca, Sr, Ba, Pb) have, in fact, shown evidence of even an upturn in the ρ vs. T curve at characteristic high temperatures ($T^*$) which have been taken as a clear indication of a reentrant metallic phase [3,4]. We refrain, at this stage, from identifying $T^*$ as $T_{JT}$ since $T^*$ is found to be dependent on other factors as well, apart from the JT effect. An understanding of occurrence of the reentrant metallic phase at high temperature is of fundamental importance because this can elucidate the role of lattice distortion in the CMR oxides, particularly at higher temperatures ($T>T_c$). At $T>T_c$, due to the presence of strong coupling of the lattice distortion to carrier hopping, the transport is governed by polaronic mechanism and $\rho$ shows an activated behavior with $\rho = \rho_0 T\exp(E_a/k_BT)$ [5]. The onset of a reentrant metallic behavior implies a change in the transport mechanism. In this letter, we report the results of our preliminary investigation that identifies the parameters that might determine the temperature $T^*$ at which the reentrant metallic phase occurs. In addition, we also investigate how the transport gap $E_a$ closes as $T \rightarrow T^*$.

*Corresponding author; e-mail: dipten@csnpl.ren.nic.in
†Currently at the Department of Physics, Indian Institute of Science, Bangalore 560 012, India

The experiments were carried out on well sintered single phase bulk pellets. A range of compositions, like, $La_{1-x}Ca_xMnO_3$ (x = 0.2-0.5), $Re_{0.5}Ae_{0.5}MnO_3$ (Re = Nd, Y and Ae = Ca and Sr) etc. has been prepared for this investigation. In addition, we utilize results of our earlier investigation [4]. The powder (of approximate size 0.5 μm) has been prepared by autoignition of citrate-nitrate gel. This method allows preparation of powder at a rather low temperature [6]. The powders thus obtained have been calcined at 1273-1373 K for 24h and finally sintered at around 1650 K for 10h in the pellet form. The phase purity, microstructure as well as compositional homogeneity of all the samples have been studied by X-ray diffraction (XRD), scanning electron microscope (SEM), and EDAX. Thermogravimetric analysis carried out till 1473 K could not detect any weight loss; 1473 K is the highest temperature attained during the resistivity measurements. The resistivities of the samples have been measured in the temperature range 15-1473 K using usual 4-probe technique. At high temperatures we used platinum paste (Make Tanaka K.K., Japan) for making the contacts with thin platinum leads. The contacts have been cured at >1000$^o$C overnight. The magnetization of the samples over 15-300 K were measured in a vibrating sample magnetometer (DMS 1660).

All the samples depicted expected resistivity and magnetization patterns within the temperature range 15-300 K. For example, the $La_{1-x}Ca_xMnO_3$ (x = 0.2-0.5) series show the usual ferromagnetic and metallic transitions ($T_c$) at temperatures in the range 200-270 K. These values of $T_c$ closely agree with those found previously in these compounds [7].

Typical resistivity data over 15-1473 K are shown in Fig. 1 for few important compositions like $La_{1-x}Ca_xMnO_3$ (x = 0.2, 0.3, 0.4, 0.45, 0.5). It can be seen that at a certain temperature $T^*$, the resistivity pattern shows an upturn and it increases as T is increased beyond $T^*$. The temperature $T^*$ is identified from the point where the derivative $d\rho/dT \geq 0$. The onset of reentrant metallic behavior and the pattern beyond are shown clearly in the inset of Fig. 1. One rather interesting point may be noted from the Fig. 1. In almost all the samples studied, the reentrant metallic transition takes place around a resistivity value 5-9 mOhm-cm. In fact, for a large class of $ABO_3$ type perovskite oxides, insulator-metal transition takes place for $\rho \approx$ 1-10 mOhm-cm [8].

The important question that has been looked into in this paper is the dependence of $T^*$ on parameters like carrier concentration and bandwidth. In Fig. 2, we show the variation of $T^*$ as a function of carrier concentration (x) for the series $La_{1-x}Ca_xMnO_3$ (LCMO) and $La_{1-x}Sr_xMnO_3$ (LSMO). We incorporate the $T_{JT}$ data of the parent compound $LaMnO_3$. And for the sake of completeness, we have also included results of earlier studies done on LCMO for x>0.5 (marked in the figure caption). It can be immediately seen that the dependence of $T^*$ on x is not monotonous and it shows a peak at x = 0.45 for LCMO and x = 0.25 for LSMO. This is an anomalous pattern. It points out that the transition temperature $T^*$ does not depend on JT effect alone. It also highlights the fact that even though one observes a sharp drop in resistivity at $T_{JT}$ in parent $LaMnO_3$, $T_{JT}$ and $T^*$ may signify two different transitions and, therefore, they may or may not coincide in a given compound. It can be mentioned, in this context, that the detailed lattice distortion and phase transition study at high temperature [1] in parent $LaMnO_3$ compound points out the presence of lattice distortion even beyond $T_{JT}$ (~750 K) and a second phase transition at around 1010 K. It will be interesting to investigate the onset of upturn in the resistivity pattern, if any, in this compound as *the upturn in the resistivity pattern at high temperature can only be observed when all types of lattice distortions are removed and the transport gap* $E_a \approx 0$. Therefore, we emphasize that $T^*$ depends on other type of lattice distortions as well apart from the JT effect. One such distortion is the mismatch in the A-site size which leads to the tilting and rotation of the $MnO_6$ octahedra and consequent modulation in the conduction bandwidth. The possibility of presence of other type of distortion and its influence on $T^*$ provides explanation to the difference in behavior between the LCMO and LSMO series. The bending of the Mn-O-Mn bonds due to tilting and rotation of $MnO_6$ octahedra is more in the former case than in the latter ($\langle r_A \rangle$ is less in the former case). Hence, the anomalous $T^*$ vs. x pattern is extended up to a higher x value in the former series. It will be interesting to study the difference and/or coincidence of $T_{JT}$ and $T^*$ for all the compositions by studying, separately, the removal of JT and other distortions through structural studies and the onset of resistivity upturn through resistivity measurements. However, for higher values of x, $T^*$ does decrease with increasing x as expected. *This result clearly shows that the carrier concentration (or the concentration of JT-active $Mn^{3+}$ ions) alone does not determine $T^*$ although it must be one of the determining factors.*

In the region $0.2<x\leq 0.5$, both LCMO and LSMO samples show a ferromagnetic transition and onset of metallic behavior at $T_c$. In the presence of double exchange interaction (DEI), the metallic state is stabilized by the ferromagnetic interaction below $T_c$. $T_c$, of course, depends on x as well as $<r_A>$. The bandwidth (BW) of the conduction band, in such a regime, is $\propto T_c$ and $T_c$ is generally taken as a measure of BW. The same samples show reentrant metallic behavior for $T\geq T^*$. The insulating state is thus stable in the temperature range $T^*$-$T_c$ (=W). As the bandwidth increases, the $T_c$ increases and the increased metallicity weakens the local JT distortion and thus destabilizes the insulating state. The enhanced bandwidth is thus expected to reduce the temperature range W, over which the insulating state is stable. In order to see a direct correlation, if any, between W and the DEI governed BW, we plot W vs. $1000/T_c$ in Fig. 3. It can be seen that W drops initially with $1000/T_c$ which implies that a smaller bandwidth compound has low $T_c$ and high $T^*$. For sufficiently small BW, the ferromagnetic state is not stable and gives way to charge and orbital order along with an antiferromagnetic order. A logical conclusion would be that for a sufficiently large bandwidth, $W\rightarrow 0$, i.e., $T^*\rightarrow T_c$. However, the W vs. $1000/T_c$ data show that W is minimum for an optimum $T_c$ and not maximum $T_c$. For even higher $T_c$, W rises again. Presence of such a minimum points out that large bandwidth (i.e. large $T_c$) alone (where the bandwidth is primarily governed by DEI) does not guarantee stability of the metallic state. The minimum in the plot is observed for a composition $La_{0.5}Sr_{0.5}MnO_3$ whereas for composition $La_{0.7}Sr_{0.3}MnO_3$ (whose $T_c$ is even higher), W is higher. The stability of the insulating state (as measured by W), therefore, is related to the concentration of JT-active $Mn^{3+}$ ions as well. Even in a compound of very high $T_c$, W can be finite if the concentration of JT ion is large. In fact, it can be observed that W drops monotonically with x in LSMO series which shows that W has one-to-one correspondence with x in such a system (see Fig. 4).

The dependence of $T^*$ on another important parameter - $<r_A>$ - can be seen, separately, if we fix x and vary $<r_A>$, a factor that is known to control the packing of the $MnO_6$ octahedron (geometric effect) and can be linked to the bandwidth. A larger $<r_A>$ ensures a larger bandwidth. In the inset of Fig. 4, we show $T^*$ as a function of the $<r_A>$ for a fixed value of x, namely, x = 0.5. We have defined $<r_A> = (r_{Re}+r_{Ae})/2$ for the composition $Re_{0.5}Ae_{0.5}MnO_3$; $r_{Re}$, $r_{Ae}$ data are taken from Shannon table [9]. It can be very clearly seen that $T^*$ drops with the rise in $<r_A>$ for a given carrier concentration.

In the regime $0.5<x\leq 1.0$, for most of the manganites, the antiferromagnetic interaction takes over and the ferromagnetic metallic state ceases to exist. Under this condition, one cannot relate the $T^*$ with bandwidth through $T_c$ any more. $T^*$, of course, is found to be dependent on the concentration of JT-active $Mn^{3+}$ ions as well as $<r_A>$ in this regime too [3]. $T^*$ drops with the increase in x and with the increase in $<r_A>$ for a fixed x.

These results point out that two factors that appear to control the $T^*$ (or high temperature insulator-metal transition), more significantly, are the carrier concentration (or the concentration of the JT-active $Mn^{3+}$ ions) and the A-site size mismatch (geometric factor) or bandwidth for a fixed x. It will be interesting to observe an interplay, if any, between these two kinds of lattice distortions – (i) JT and (ii) geometric. From a detailed local structural distortion study, one can unravel the extent of interplay, if any. This, however, is beyond the scope of this present work.

The reentry into the metallic phase at $T>T^*$ implies that the transport gap closes at $T\geq T^*$. We investigate this by evaluating the temperature dependence of the transport activation energy $E_a$. In Fig. 5, we plot $d\ln(\rho/T)/d(1/T)$ as a function of T. In the temperature range $T_c<T<T^*$, the resistivity follows the relation $\rho = (\rho_0 T).\exp(E_a/k_B T)$ as expected in a polaronic transport scenario [5]. One would expect that the derivative $d\ln(\rho/T)/d(1/T)$ gives a measure of the activation energy $E_a$ at a temperature T. Recently, it has been seen that at the approach of $T_c$ from higher temperature there is a significant deviation from the polaronic behavior [10]. In this paper, of course, our concern is the region close to $T^*$ where the transport gap is expected to close. It can be seen that there are indeed two distinct cases. In the compounds $La_{1-x}Ca_xMnO_3$ (x = 0.4-0.5), the collapse of the gap is rather sharp. $E_a$ is essentially temperature independent for most of the temperature region, but it closes to zero rather sharply as $T\rightarrow T^*$. For other compositions, the gap $E_a$ approaches zero gradually. This may be the result of inhomogeneity in the composition across the matrix which gives rise to a broader transition.

These results are important in the context of estimation of role of lattice distortion in governing the entire range of behaviors – CMR, CO, magnetic field driven insulator to metal transition etc. – in this class of materials. From our study, in which we show high temperature insulator to metal transition in a range of compounds, it is quite clear that it may be possible to estimate the extent of effective lattice distortion from a measurement of $T^*$. From the knowledge of the relevant factors which influence the $T^*$, it will be possible to engineer the effective distortion subtly for a desired behavior as the structural, transport and magnetic properties are intimately related in these compounds.

In summary, we provide data on insulator to metal transition (and onset of lattice distortion) at higher temperature (beyond $T_c$) in a series of rare-earth manganites. We have also pointed out that the transition temperature ($T^*$) depends on the carrier concentration (x) or the concentration of Jahn-Teller-active $Mn^{3+}$ ions, the double exchange interaction governed bandwidth as well as on the geometric factor (A-site size mismatch) for compounds with fixed x. However, in the absence of a detailed structural study, it may not always be possible to separate out the influence of any particular factor. Our study underscores the need to carry out such high temperature measurements in order to estimate the effective lattice distortion which governs the CMR, CO effects etc. significantly. It will be interesting to study any interplay between the Jahn-Teller effect and the geometric effect through estimation of the nature and extent of local lattice distortion. This issue will be addressed in a future work.

Authors acknowledge thankfully the assistance rendered by A.K. Halder and S.K. Pratihar (CGCRI) during the high temperature resistivity measurements. They also acknowledge support of R.G. Sharma (NPL) and H.S. Maiti (CGCRI). One of authors' (AKR) acknowledges the financial support from BRNS and two other authors (P.D. and A.P.) acknowledge financial support from UGC and CSIR, Govt. of India.

**Figure Captions**

Fig.1. The resistivity vs. temperature plots over a regime 15-1473 K for some of the important compounds: $La_{0.8}Ca_{0.2}MnO_3$ (●), $La_{0.7}Ca_{0.3}MnO_3$ (■), $La_{0.6}Ca_{0.4}MnO_3$ (▲), $La_{0.55}Ca_{0.45}MnO_3$ (▼), and $La_{0.5}Ca_{0.5}MnO_3$ (◆). Inset : the onset of insulator to metal transition around $T^*$ and the resistivity pattern beyond $T^*$ are shown.

Fig. 2. Variation of $T^*$ with carrier concentration (x) for $La_{1-x}Sr_xMnO_3$ (■) and $La_{1-x}Ca_xMnO_3$ (●) series. The data for LCMO series in the range x>0.5 are taken from Ref. [3].

Fig.3. Variation of W ($=T^*-T_c$) with $1000/T_c$ in the compounds depicting both the ferromagnetic metallic to paramagnetic insulating transition as well as high temperature insulator to metal transition.

Fig.4. Variation of W ($=T^*-T_c$) as a function of carrier concentration (x) in $La_{1-x}Sr_xMnO_3$ series; inset shows the variation of $T^*$ as a function of average A-site radius $\langle r_A \rangle$ for a fixed carried concentration x = 0.5.

Fig.5. The dependence of activation energy $E_a$ [$d\ln(\rho/T)/d(1/T)$] on temperature. The drop of $E_a$ to zero marks the onset of reentrant metallic behavior. The nature of drop, of course, varies from compound to compound.

**Fig.1. D. Bhattacharya *et al.***

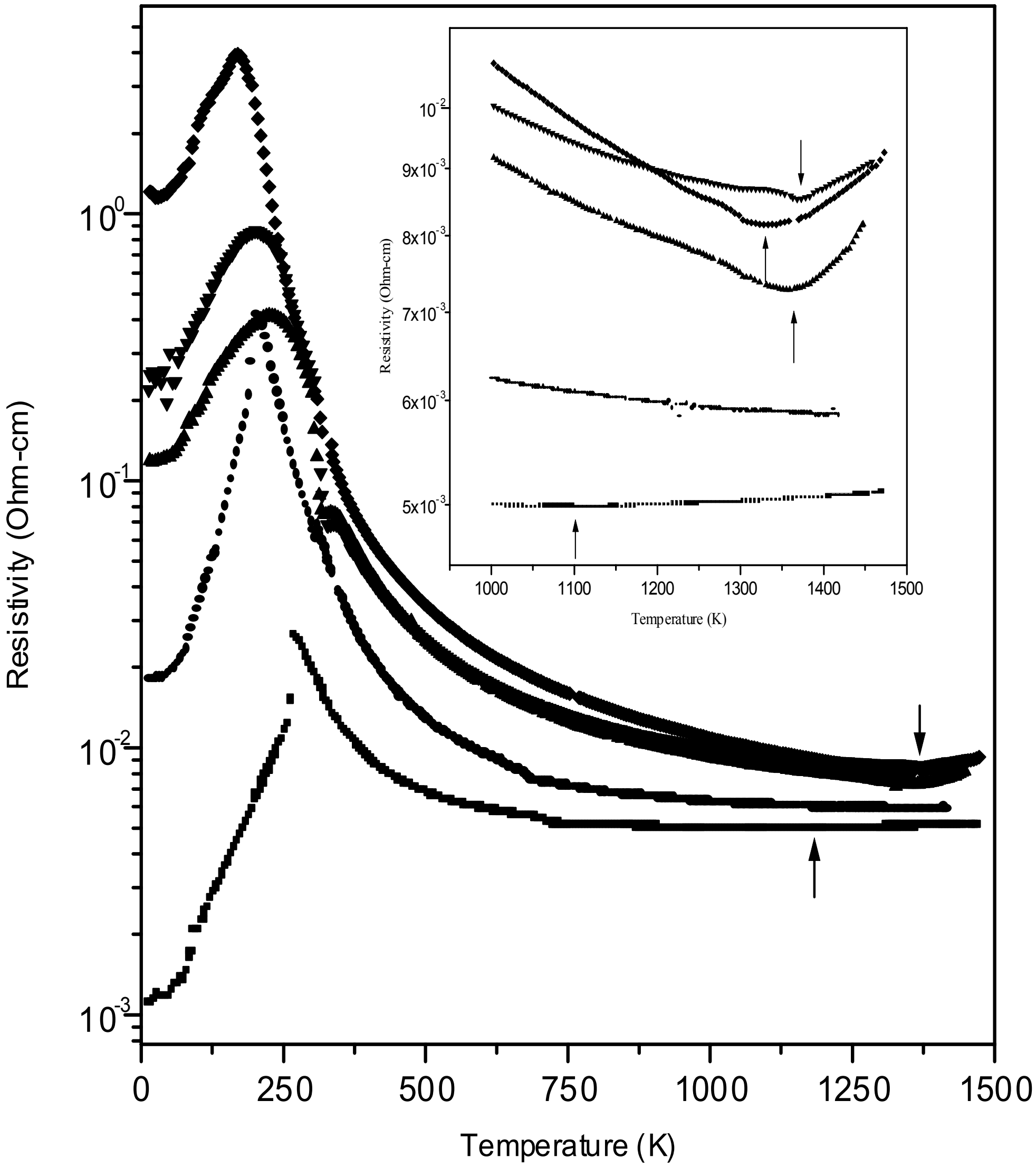

**Fig.2. D. Bhattacharya *et al.***

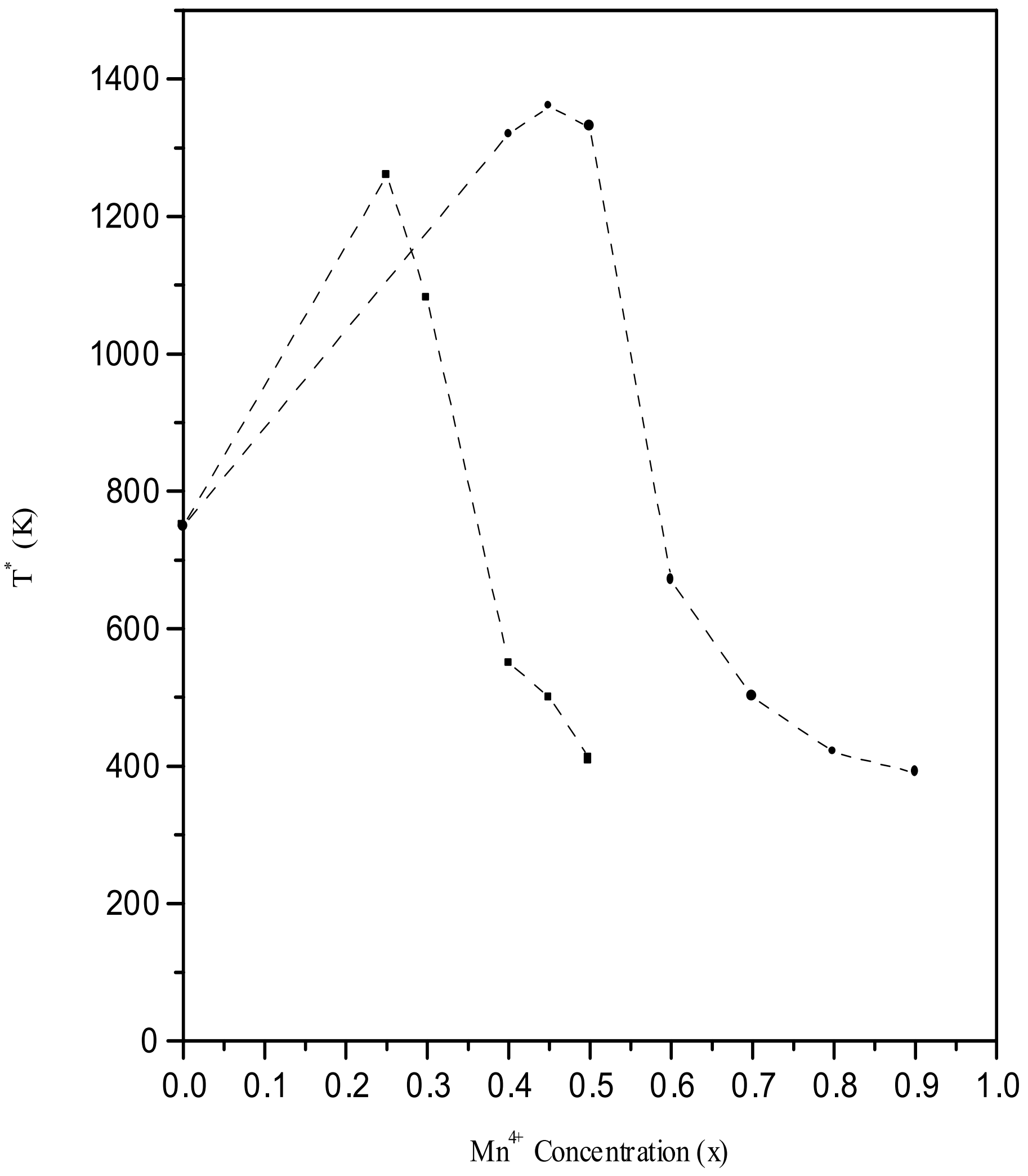

**Fig.3. D. Bhattacharya *et al.***

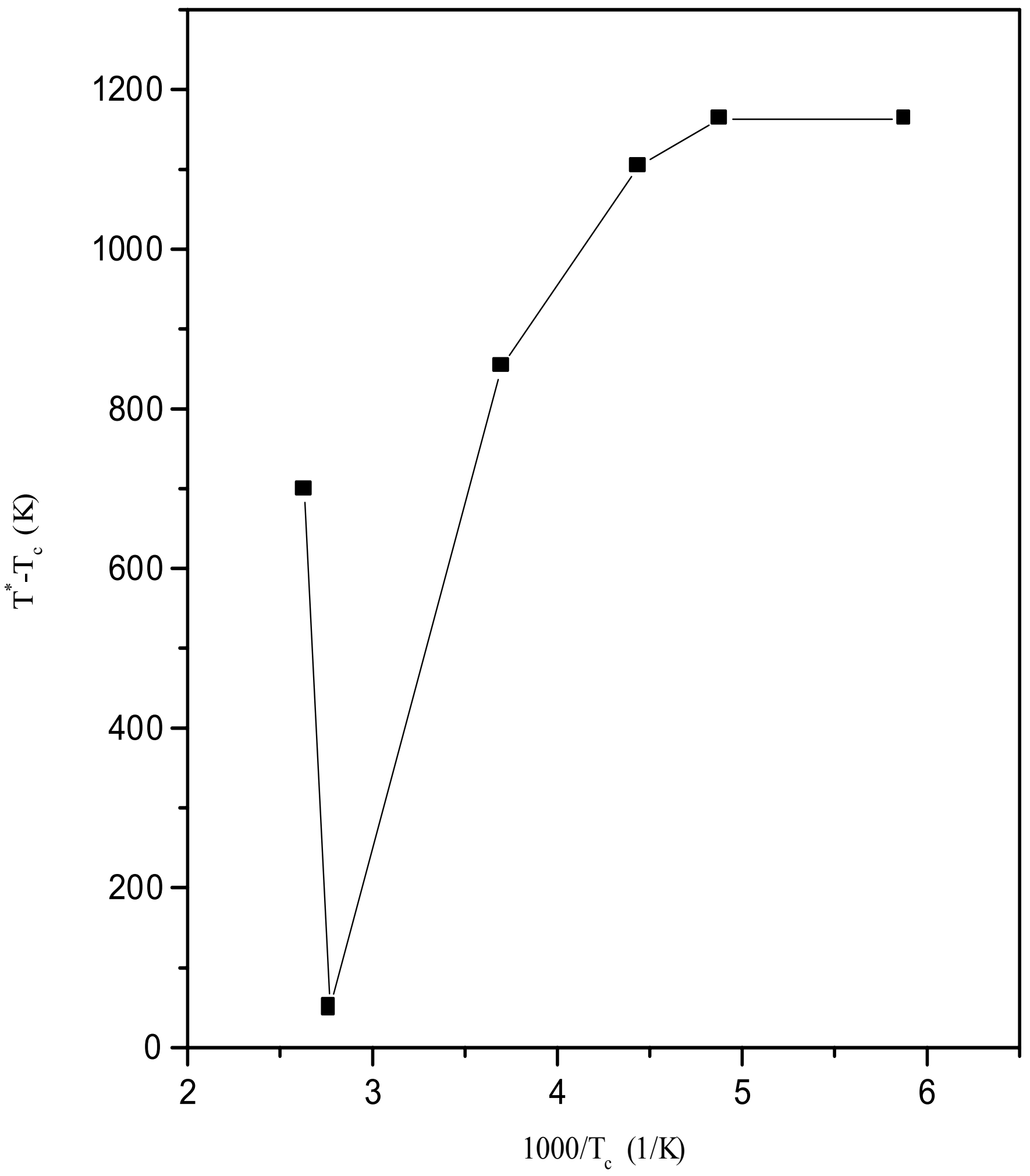

**Fig.4. D. Bhattacharya *et al.***

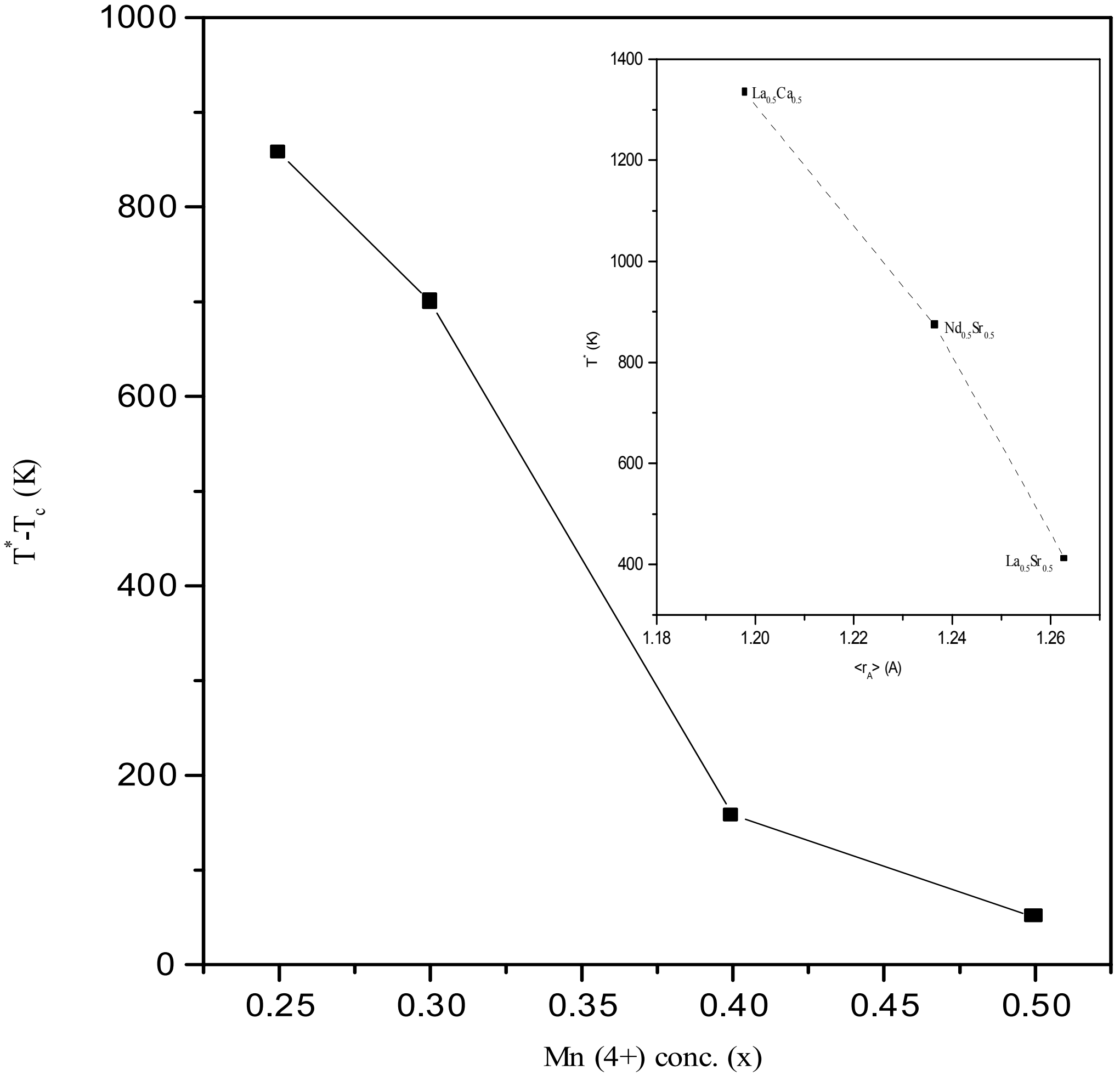

**Fig.5. D. Bhattacharya *et al.***

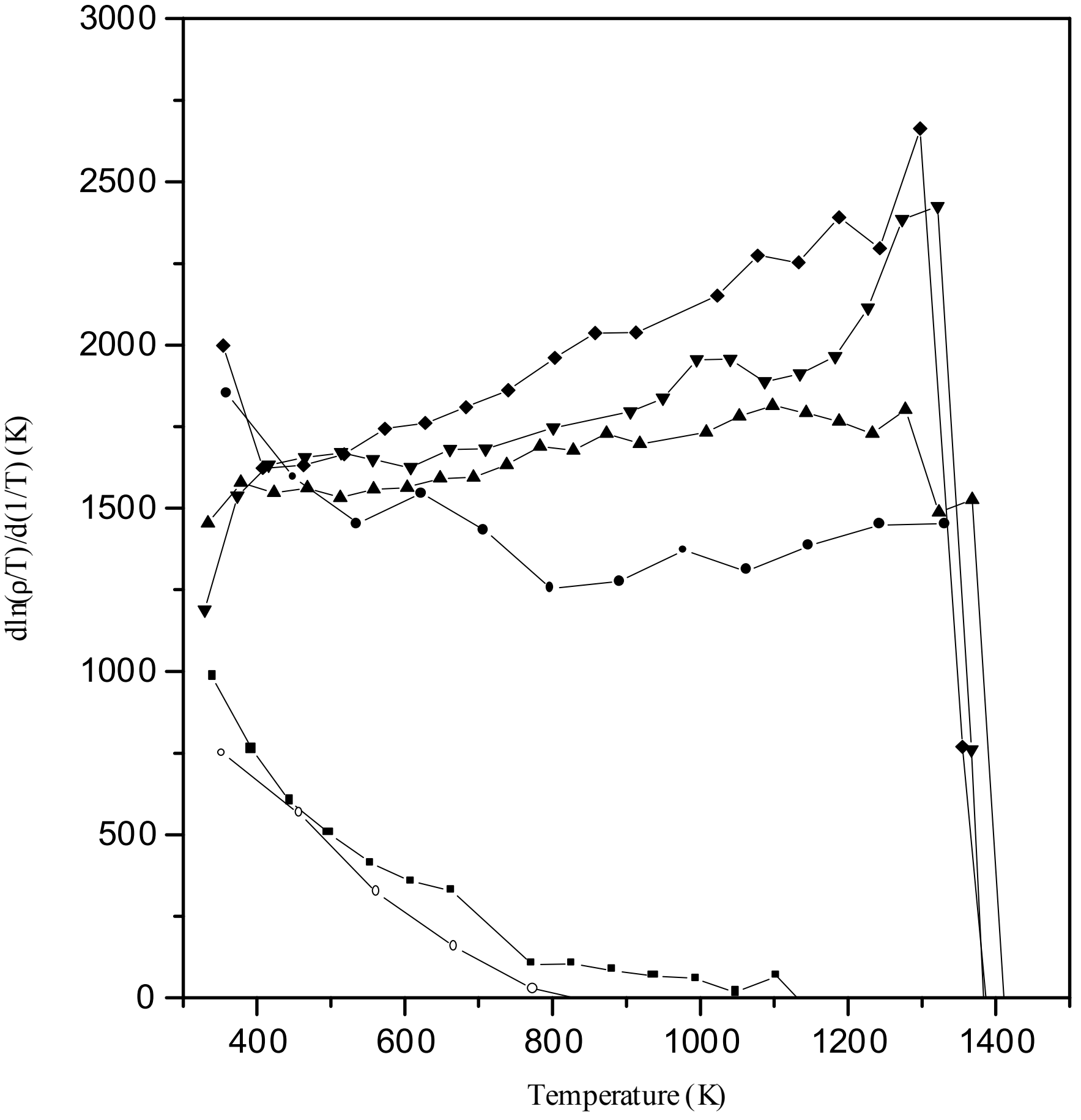